\documentclass[aps,prl,groupedaddress,amsmath,amssymb,tightenlines,notitlepage,footinbib,twocolumn]{revtex4-1}
\usepackage{graphicx}
\usepackage[dvipsnames,svgnames]{xcolor}
\usepackage{bm,bbm}%
\usepackage{braket}
\usepackage{amsthm,amsmath,amssymb,amsbsy}
\usepackage{enumerate}
\usepackage[caption=false]{subfig}
\usepackage{booktabs,multirow,moresize}
\usepackage{mathtools,mathrsfs,bbding}
\usepackage[percent]{overpic}
\usepackage{microtype}

\usepackage{chngcntr}

\usepackage[T1]{fontenc}
\usepackage[utf8]{inputenc}

\definecolor{darkred}{rgb}{0.57,0,0.12}
\usepackage[colorlinks=true, linkcolor=MidnightBlue, urlcolor=MidnightBlue, citecolor=MidnightBlue]{hyperref}

\usepackage{cleveref}

\let\nc\newcommand

\usepackage{newpxtext,newpxmath}

\hypersetup{
    bookmarksnumbered=true, 
    breaklinks=true,
    unicode=false, 
    pdfstartview={FitH}, 
    pdftitle={}, 
    pdfnewwindow=true, 
    colorlinks=true, 
    allcolors=MidnightBlue!70!black!70!TealBlue
}

\nc{\note}[1]{{\color{blue!90!black} #1}}
\nc{\noteb}[1]{{\color{red!80!black}\textbf{#1}}}

\nc{\incoh}[1]{{\mathcal{C}^{(#1-1)}}}
\nc{\produc}[1]{{\mathcal{P}^{(#1)}}}

\nc{\cmin}{c_{\min}}
\nc{\CG}{C_{\text{GME}}}

\nc{\gram}[1]{G^{(#1)}}

\nc{\pen}{^{-1}}

\DeclareMathOperator{\Tr}{Tr}

\DeclareMathOperator{\supp}{supp}

\DeclareMathOperator{\aff}{aff}

\nc{\psic}{\psi^{c}}
\nc{\two}[1]{\underline{2^{d-#1}}}

\nc{\Hanc}{\mathcal{H}_{\text{anc}}}
\nc{\psianc}{\psi_{\text{anc}}}
\nc{\lampow}{\lambda^{1/d}}

\nc{\norm}[2]{\left\lVert#1\right\rVert_{\,#2}}
\nc{\proj}[1]{\ket{#1}\!\bra{#1}}
\nc{\pro}[1]{#1 #1^\dagger}
\nc{\lnorm}[2]{\left\lVert#1\right\rVert_{\ell_{#2}}}

\nc{\VV}{V^{\,\C}_\Sp}

\nc{\R}{\mathcal{R}}
\nc{\W}{\mathcal{W}}
\nc{\T}{\mathcal{T}}
\nc{\B}{\mathcal{B}}
\nc{\C}{\mathcal{C}}
\nc{\U}{\mathcal{U}}
\nc{\E}{\mathcal{E}}
\nc{\EE}{\mathscr{E}}
\nc{\K}{\mathcal{K}}
\nc{\V}{\mathcal{V}}
\nc{\X}{\mathcal{X}}
\nc{\F}{\mathbb{F}}
\nc{\D}{\mathcal{D}}
\nc{\Y}{\mathcal{Y}}

\nc{\M}{\mathcal{M}}
\nc{\N}{\mathcal{N}}
\nc{\I}{\mathcal{I}}
\nc{\Q}{\mathbb{Q}}
\nc{\RR}{\mathbb{R}}
\nc{\CC}{\mathbb{C}}

\nc{\HH}{\mathbb{H}}
\nc{\MM}{\mathbb{M}}
\nc{\DD}{\mathbb{D}}

\nc{\J}{\mathcal{J}}

\let\FF\F
\let\OO\O

\let\succc\succ

\nc{\mleq}{\preceq}
\nc{\mgeq}{\succeq}
\nc{\mle}{\prec}
\nc{\mge}{\succc}

\renewcommand{\succ}{{\mathrm{succ}}}

\nc{\CPTP}{{\mathrm{CPTP}}}
\nc{\CP}{{\mathrm{CP}}}
\nc{\CPTN}{{\mathrm{CPTN}}}

\nc{\ext}[1]{\operatorname{ext}\left(#1\right)}

\nc{\<}{\left\langle}

\renewcommand{\bar}{\;\rule{0pt}{9.5pt}\right|\;}

\nc{\lset}{\left\{\left.}
\nc{\rset}{\right\}}

\DeclareMathOperator{\cone}{cone}

\nc{\ve}{\varepsilon}

\nc{\cbraket}[1]{\left|\braket{#1}\right|}

\nc{\id}{\mathbbm{1}}
\nc{\idc}{\mathrm{id}}

\nc{\mnorm}[1]{\norm{#1}{[m]}}
\nc{\knorm}[1]{\norm{#1}{(k)}}

\theoremstyle{plain}

\newtheorem{theorem}{Theorem}

\newtheorem{corollary}[theorem]{Corollary}

\theoremstyle{definition}

\let\oldproofname\proofname
\renewcommand{\proofname}{\rm\bf{\oldproofname}}

\nc{\lsetr}{\left\{\,}
\nc{\rsetr}{\right.\right\}}
\nc{\barr}{\;\rule{0pt}{9.5pt}\left|\;}
\nc{\ketbra}[2]{\ket{#1}\!\bra{#2}}

\nc{\prob}{\mathrm{prob}}

\nc{\eqt}[1]{\stackrel{\mathclap{\mbox{\scriptsize #1}}}{=}}

\nc{\wt}{\widetilde}

\nc{\Rmax}{R_{\max}}
\nc{\Pmax}{R_{\max}}
\nc{\Qmin}{Q_{\min}}
\nc{\Qmax}{V_{\FF}}
\nc{\Qmaxa}{V_{\aff(\FF)}}
\nc{\Rs}{R_{s}}

\nc{\RW}{\Omega_\F}

\nc{\din}{d_\mathrm{in}}
\nc{\dout}{d_\mathrm{out}}

\nc{\gbar}{,\;}

\nc{\Hh}{\wt{H}}

\nc{\HP}{\Xi}

\nc{\bb}{{\bullet\bullet}}
\nc{\phim}{{\phi_{\star}}}

 \usepackage{titlesec}
 \titlespacing*{\subsection}{0pt}{1.1\baselineskip}{\baselineskip}

 \nc{\ba}{\begin{equation}\begin{aligned}}
\nc{\ea}{\end{aligned}\end{equation}}

\begin{document}

 \title{Probabilistic transformations of quantum resources}

 \author{Bartosz Regula}
 \email{bartosz.regula@gmail.com}
\affiliation{Department of Physics, Graduate School of Science, The University of Tokyo, Bunkyo-ku, Tokyo 113-0033, Japan}

\begin{abstract}
The difficulty in manipulating quantum resources deterministically often necessitates the use of probabilistic protocols, but the characterization of their capabilities and limitations has been lacking. We develop a general approach to this problem by introducing a new resource monotone that obeys a very strong type of monotonicity: it can rule out all transformations, probabilistic or deterministic, between states in any quantum resource theory. 
This allows us to place fundamental limitations on state transformations and restrict the advantages that probabilistic protocols can provide over deterministic ones, significantly strengthening previous findings and extending recent no-go theorems.
We apply our results to obtain a substantial improvement in bounds for the errors and overheads of probabilistic distillation protocols, directly applicable to tasks such as entanglement or magic state distillation, and computable through convex optimization. 
In broad classes of resources, we strengthen our results to show that the monotone completely governs probabilistic transformations --- its monotonicity provides a necessary and sufficient condition for state convertibility. This endows the monotone with a direct operational interpretation, as it can exactly quantify the highest fidelity achievable in resource distillation tasks by means of any probabilistic manipulation protocol.
\end{abstract}

 \maketitle

The extent of our ability to manipulate different quantum resources determines how well we can utilize them in practice. This can be broadly formulated as the question of resource convertibility: when can one resource state be transformed into another under the restrictions imposed by a given physical setting? A representative example of such conversion is the process of distillation (purification), indispensable in the applications of resources such as quantum entanglement~\cite{bennett_1996} and magic states~\cite{bravyi_2005} due to the noise inherent in the preparation and manipulation of quantum systems. Understanding the exact conditions for the existence of physical transformations between quantum states has attracted significant attention from the early days of quantum information theory~\cite{alberti_1980,nielsen_1999,vidal_1999-1,chefles_2004,buscemi_2012,reeb_2011,heinosaari_2012,horodecki_2013,alhambra_2016,buscemi_2017,gour_2017,gour_2018-2,takagi_2019,liu_2019,buscemi_2019,dallarno_2020,regula_2020,zhou_2020}.
In addition to providing insight into practical resource manipulation protocols, such results establish the ultimate limits of resource manipulation in the form of no-go theorems that identify the regimes in which certain transformations are not simply difficult to perform, but truly physically impossible.
This latter aspect is particularly important as it allows one to certify the optimality of practical protocols and reveal fundamental restrictions imposed by the laws of quantum mechanics.

The formalism of quantum resource theories~\cite{horodecki_2013-3,chitambar_2019} provides the tools to understand and constrain the conversion of resources in a unified approach across different physical settings. An important part of such toolsets are resource monotones, which can establish restrictions on feasible conversion schemes. Although this avenue has lately seen significant attention in the context of general resource theories~\cite{brandao_2015,gour_2017,takagi_2019,liu_2019,regula_2020,fang_2020,gonda_2019,hsieh_2020,kuroiwa_2020,ferrari_2020,regula_2021-1,fang_2020-2}, it has several limitations.
First, deciding the convertibility between states in general settings typically requires one to compare infinitely many monotones~\cite{buscemi_2012,gour_2017,takagi_2019,gour_2020-1}, hindering the practical applicability of these methods.
Another major downside to many of such approaches is that they can only characterize deterministic transformations, i.e., ones which succeed with certainty. Due to the difficulty in realizing such exact transformations, practical protocols typically exploit measurement-based schemes that are inherently probabilistic in nature~\cite{horodecki_2009,campbell_2017}. It is already known from the theory of quantum entanglement that probabilistic manipulation methods, such as stochastic local operations and classical communication (SLOCC), can significantly enhance our capability to perform certain transformations~\cite{lo_2001,vidal_1999-1,dur_2000,horodecki_1999-1}. It then becomes an important problem to extend general resource-theoretic approaches to completely characterize also non-deterministic resource manipulation.

To address this, we introduce a new resource monotone, the projective robustness $\RW$, which obeys the strongest type of monotonicity: it can never increase under any resource transformation, deterministic or probabilistic. The measure is computable through convex optimization, providing an accessible criterion that can rule out \emph{all} transformations between pairs of states, including the most general forms of probabilistic protocols. We use this to significantly strengthen and extend previously known no-go theorems in the probabilistic manipulation of resources. In particular, we provide tighter bounds on the errors and overheads incurred in probabilistic distillation tasks, revealing fidelity thresholds that cannot be achieved through any physical means.
Our results establish the ultimate limitations on the performance of all resource manipulation protocols, thus also constraining the advantages that can be gained by employing probabilistic transformations over deterministic ones. The methods rely only on the basic laws of quantum mechanics and are therefore directly applicable in a wide variety of physical contexts, which we exemplify through applications to concrete settings such as entanglement, coherence, and magic state distillation.

Notably, our no-go theorems can serve not only as necessary conditions for resource manipulation, but also as sufficient ones. We show that in several types of quantum resources --- e.g.\ in the whole class of affine resource theories, which includes important examples such as coherence, asymmetry, and imaginarity, as well as in the distillation of quantum entanglement with non-entangling operations --- the projective robustness $\RW$ completely governs probabilistic convertibility between states. This establishes $\RW$ as an operationally meaningful monotone that plays a fundamental role in understanding the manipulation of quantum resources, exactly quantifying the best achievable fidelity in resource distillation achievable through any probabilistic protocol.

Here we introduce the main concepts and results. The detailed technical derivations, as well as additional details and extensions can be found in~\cite{regula_2021-4}.

\subsection*{Resource transformations}
Resource theories are frameworks concerned with manipulating quantum systems under some physical constraints on the allowed states and operations~\cite{horodecki_2013-3,chitambar_2019}. Any such restriction singles out the \emph{free states} $\FF$ and the \emph{free operations} $\OO$ that are allowed within the restricted setting. For our results to be as general as possible, we only make two basic assumptions about the set of free states $\FF$: that it is closed and convex. In a similar axiomatic manner, we take $\OO$ to be the maximal physically consistent set of free operations, namely, all channels that do not generate a given resource. Deterministic transformations $\E \in \OO$ are then completely positive and trace-preserving maps for which $\E(\sigma) \in \FF$ for all free $\sigma$. 

We will model probabilistic transformations using stochastic quantum operations, that is, ones that are not necessarily trace preserving. Such maps can be thought of as being part of a quantum instrument~\cite{davies_1970,ozawa_1984} composed of free operations. 
The question of probabilistic convertibility between two states then reduces to the existence of a map $\E(\rho) = p \rho'$ for some $p \in (0,1]$ where $\E$ is a sub-normalized free operation, that is, a completely positive and trace--non-increasing map such that $\E(\sigma) \propto \sigma' \in \FF$ for all $\sigma \in \FF$. We use $\OO$ to denote both deterministic and probabilistic (trace--non-increasing) free maps.
In the settings of interest discussed in this work, this is the largest possible choice of free probabilistic transformations, meaning that all no-go results shown for the operations $\OO$ will necessarily apply to any other physical type of free operations. 

\subsection*{Resource monotones}
A resource monotone $M_\FF$ is any function which is monotonic under deterministic free operations, that is, $M_\FF(\rho) \geq M_\FF(\E(\rho))$ for a channel $\E \in \OO$~\cite{vidal_2000,chitambar_2019}. Any such monotone can be used to rule out the existence of a transformation between two states --- if $M_\FF(\rho) < M_\FF(\rho')$, then there cannot exist a free operation that transforms $\rho$ to $\rho'$ with certainty. The situation is more complicated when probabilistic transformations are concerned~\cite{vidal_2000}, and in fact most known resource monotones can never be used to rule out the existence of such stochastic protocols.

Two monotones that have found a variety of uses in the description of quantum resources are the \emph{(generalized) robustness}~$R_\FF$~\cite{vidal_1999,datta_2009} and the \emph{resource weight}~$W_\FF$~\cite{lewenstein_1998}. The two measures can both be expressed in terms of the max-relative entropy $D_{\max}$~\cite{datta_2009}, the non-logarithmic variant of which we define as
\ba
  \Rmax(\rho \| \sigma) \coloneqq 2^{D_{\max}(\rho\|\sigma)} = \inf \lset \lambda \bar \rho \leq \lambda \sigma \rset,
\ea
where the inequality $\rho \leq \lambda \sigma$ means that $\lambda \sigma - \rho$ is positive semidefinite.
We can then write $R_\FF(\rho) \coloneqq \min_{\sigma \in \FF} \Rmax(\rho \| \sigma)$ and $W_\FF(\rho) \coloneqq [\min_{\sigma \in \FF} \Rmax(\sigma \| \rho)]^{-1}$.

Although both the robustness and weight are operationally useful monotones~\cite{takagi_2019-2,regula_2020,uola_2020-1,ducuara_2020,haapasalo_2014,regula_2021-1,fang_2020-2}, they generally do not provide tight restrictions on probabilistic transformations~\cite{regula_2021-1,fang_2020-2}. The starting point of this Letter is the observation that stronger limitations can be obtained by introducing a new monotone which combines the properties of the two.


\subsection*{Projective robustness}
We define the \textit{projective robustness} $\RW$ as
\ba
  \RW(\rho) \coloneqq \min_{\sigma \in \FF}\, \Rmax(\rho \| \sigma) \, \Rmax(\sigma \| \rho).
\ea
The quantity owes its name to the fact that, for fixed $\rho$ and $\sigma$, the expression $\ln \Rmax(\rho \| \sigma) \, \Rmax(\sigma \| \rho)$ is known as the Hilbert projective metric~\cite{bushell_1973,kohlberg_1982,reeb_2011}. This metric has found use in understanding transformations of pairs of quantum states~\cite{reeb_2011,buscemi_2017}, but has not been applied as a resource monotone before.

The projective robustness obeys a number of useful properties~\cite{regula_2021-4}. It is important to note that $\RW$ will not be finite for all states: $\RW(\rho) < \infty$ if and only if there exists a free state $\sigma$ such that $\supp \rho = \supp \sigma$. In most commonly encountered resource theories, this means that $\RW$ will always be finite for full-rank states, but will diverge to infinity for resourceful pure states. However, since adding a small amount of noise can make any state full rank, we will see that it can lead to useful results even when pure states are considered.

Importantly, $\RW$ can always be computed as a convex optimization problem. We have that
\ba
  \RW(\rho) &= \inf \lset \gamma \in \RR_+ \bar \rho \leq \wt\sigma \leq \gamma \rho,\; \wt\sigma \in \cone(\FF) \rset\\
  &= \sup \lset \frac{\Tr A \rho }{\Tr B \rho } \bar A, B \geq 0,\, \frac{\Tr A \vphantom{\rho}\sigma}{\Tr B \vphantom{\rho}\sigma} \leq 1 \; \forall \sigma \in \FF \rset,
\ea
where $\cone(\FF) = \lset \lambda \sigma \bar \lambda \in \RR_+, \; \sigma \in \FF \rset$ denotes the cone generated by the set of free states. In many relevant cases of resources--- such as quantum coherence, magic, or the theory of non-positive partial transpose --- this optimization reduces to a semidefinite program.

\subsection*{No-go theorem for resource transformations}
A notable property of the projective robustness is the very strong form of monotonicity that it obeys: $\RW$ cannot be increased by any free operation, not even probabilistically.

\begin{theorem}\label{thm:nogo_monotonicity}
If there exists a free transformation from $\rho$ to $\rho'$, probabilistic or deterministic, then $\RW(\rho) \geq \RW(\rho')$.
\end{theorem}
The Theorem thus establishes a necessary condition for the existence of any probabilistic transformation between two states, valid in all quantum resource theories.

At the basic level, the result shows that whenever there exists a free probabilistic operation $\E \in \OO$ which transforms $\rho$ to some output state with non-zero probability, it necessarily holds that $\RW(\rho) \geq \RW(\E(\rho)) =  \RW\left(\frac{\E(\rho)}{\Tr \E(\rho)}\right)$. However, the statement of Theorem~\ref{thm:nogo_monotonicity} is actually stronger than this. An intriguing phenomenon in probabilistic resource manipulation is that there exist cases where the transformation $\rho \to \rho'$ is impossible with any non-zero probability, but one can nevertheless approach $\rho'$ arbitrarily closely~\cite{horodecki_1999-1}. Then, there can exist a sequence of operations $(\E_n)_{n} \in \OO$ such that $\Tr \E_n(\rho) \to 0$ but $\frac{\E_n(\rho)}{\Tr \E_n(\rho)} \to \rho'$.
That is, the transformation might only be possible asymptotically, with probability of success vanishing as the fidelity approaches 1. The monotonicity of $\RW$ covers also this case: whenever such a protocol exists, we must have $\RW(\rho) \geq \RW(\rho')$. Put another way: whenever $\RW(\rho) < \RW(\rho')$, there cannot exist any physical free transformation taking $\rho$ arbitrarily close to $\rho'$.

In addition to providing quantitative limitations that we will explore shortly, the restriction imposed here immediately strengthens and generalizes the no-go theorems of Ref.~\cite{fang_2020} --- no state with a finite $\RW$ can be transformed to a state $\rho'$ for which $\RW(\rho') = \infty$, where the latter includes all pure resourceful states. This also extends previous results that dealt with the impossibility of entanglement~\cite{kent_1998,jane_2002,horodecki_1999-1,horodecki_2006-1,regula_2019-2} and coherence purification~\cite{fang_2018,wu_2020}.

Another monotone with similar type of monotonicity is the Schmidt number in entanglement theory~\cite{terhal_2000}, but this quantity is significantly more difficult to compute than $\RW$ for general quantum states, and its discrete character makes it unclear whether it could lead to tight quantitative bounds like the ones that we consider in this work.

\subsection*{Sufficient condition for probabilistic transformations}
We will now show that Theorem~\ref{thm:nogo_monotonicity} can also give a sufficient condition for transforming resources. In such cases, it follows that the projective robustness $\RW$ completely governs the ability to transform one state into another through probabilistic means, as long as all resource--non-generating operations $\OO$ are considered~\footnote{In Theorems 2 and 3, we make a minor technical assumption that $R_\FF(\rho') < \infty$; this is always satisfied in generally encountered resource theories, and we only assume this to avoid pathological cases, which are treated in~\cite{regula_2021-4} for completeness.}.
This result will require us to consider different types of resource theories separately.

We begin with an important class of theories dubbed \emph{affine resource theories}~\cite{gour_2017}, which includes quantum coherence~\cite{baumgratz_2014}, asymmetry~\cite{gour_2008}, thermodynamics~\cite{horodecki_2013}, or the recently studied imaginarity~\cite{hickey_2018,wu_2021}. Such theories are distinguished by the fact that the set of free states $\FF$ is the intersection of some affine subspace with the set of all states.

\begin{theorem}\label{thm:nogo_affine}
In any affine resource theory, there exists a free probabilistic transformation $\rho \to \rho'$ if and only if $\RW(\rho) \geq \RW(\rho')$.
\end{theorem}
The proof of this result combines insights from the characterization of affine resource measures in Ref.~\cite{regula_2020} with a proof method of Ref.~\cite{reeb_2011}, where transformations of pairs of quantum states were considered.

The case of non-affine theories requires a slightly different approach. Let us introduce a variant of the max-relative entropy $\Rmax$ as
\ba
  \Rmax^\FF (\rho \| \sigma) \coloneqq \inf \lset \lambda \bar \rho \leq_{\FF} \lambda \sigma\rset,
\ea
where $A \leq_\FF B \iff B - A \in \cone(\FF)$. The \emph{standard robustness} $R^\FF_\FF$~\cite{vidal_1999}, also known as free robustness, is a resource monotone defined as $R^\FF_\FF(\rho) \coloneqq \min_{\sigma \in \FF} \Rmax^\FF (\rho \| \sigma)$. We analogously define the \textit{{free projective robustness}} as
\ba\label{eq:proj_free_def}
  \RW^\FF(\rho) \coloneqq \min_{\sigma \in \FF}\, R_{\max}^\FF(\rho \| \sigma) \, \Rmax(\sigma \| \rho).
\ea
Notice that $\RW^\FF(\rho) \geq \RW(\rho)$ in general. 
$\RW^\FF$ is not useful in affine resource theories as it diverges for all resourceful states; it can, however, be applied in theories where the set of free states $\FF$ is of full measure in the set of all states.
Such theories are known as \emph{full dimensional}, and include e.g.\ quantum entanglement~\cite{horodecki_2009} or magic (non-stabilizerness)~\cite{veitch_2014,howard_2017}. We then give the following sufficient condition, which we will see to be necessary and sufficient in relevant cases.
\begin{theorem}\label{thm:nogo_sufficient_fulldim}
In any full-dimensional resource theory, there exists a free probabilistic transformation $\rho \to \rho'$ if $\RW(\rho) \geq \RW^\FF(\rho')$. Conversely, when such a transformation exists, then $\RW(\rho) \geq \RW(\rho')$ and $\RW^\FF(\rho) \geq \RW^\FF(\rho')$.
\end{theorem}

\subsection*{Probabilistic resource distillation}
The task of distillation is concerned with extracting some noiseless, pure resource state $\phi$ from a noisy state $\rho$. Since such transformations are often very difficult to achieve exactly, we allow for a small error in the conversion. We are then concerned with achieving a transformation $\rho \to \tau$ probabilistically, where $\tau$ is a state close to the target state in fidelity: $F(\tau, \phi) \geq 1-\ve$. Our aim will be to establish thresholds on the fidelity achievable through any such protocol.

First studied in entanglement theory~\cite{lo_2001,linden_1998,kent_1998,horodecki_1999-1,vidal_2000-1,horodecki_2006-1,rozpedek_2018}, purification tasks of this type have recently been investigated in the context of general resource theories. Of particular relevance are recent bounds based on the robustness $R_\FF$~\cite{regula_2020,regula_2021-1} and the resource weight $W_\FF$~\cite{regula_2021-1,fang_2020-2}, although most of their applications were limited to deterministic protocols. In a different approach, Ref.~\cite{fang_2020} showed a relation that we will refer to as the eigenvalue bound, aiming to understand trade-offs between the error $\ve$ and the achievable probability $p$ of successful transformation. However, we can show~\cite{regula_2021-4} that, as a restriction on achievable probabilities, the bound of Ref.~\cite{fang_2020} is vacuous: its dependence on the probability $p$ is superficial, and the bound holds equally for all probabilistic protocols. We will also see that, as a bound on distillation error, it performs much worse than restrictions obtained using $\RW$.

We now use the projective robustness to establish a general error threshold that cannot be exceeded by any probabilistic protocol in any quantum resource theory.
\begin{theorem}\label{thm:prob_error}
If there exists a free transformation $\rho~\to~\tau$ such that $\tau$ is a state satisfying $F(\tau,\phi) \geq 1-\ve$ for some resourceful pure state $\phi$, then
\ba
  \ve \geq \left( \frac{F_\FF(\phi)}{1-F_\FF(\phi)} \,\RW(\rho)\, + 1 \right)^{-1}
\ea
where $F_\FF(\phi) = \max_{\sigma \in \FF} \braket{\phi|\sigma|\phi}$.
\end{theorem}

Such probabilistic distillation thresholds have previously attracted significant attention in the study of quantum entanglement~\cite{linden_1998,kent_1998,horodecki_1999-1,horodecki_2006-1}, but even in that case no general quantitative bound was known. In entanglement theory, the maps $\OO$ correspond to all non-entangling protocols, and the result can be thought of as a threshold for the manipulation of entanglement under such extended transformations, or under the more restricted class of LOCC.

\begin{figure}[t]
\centering
\hspace{-0.05cm}\includegraphics[width=8.5cm]{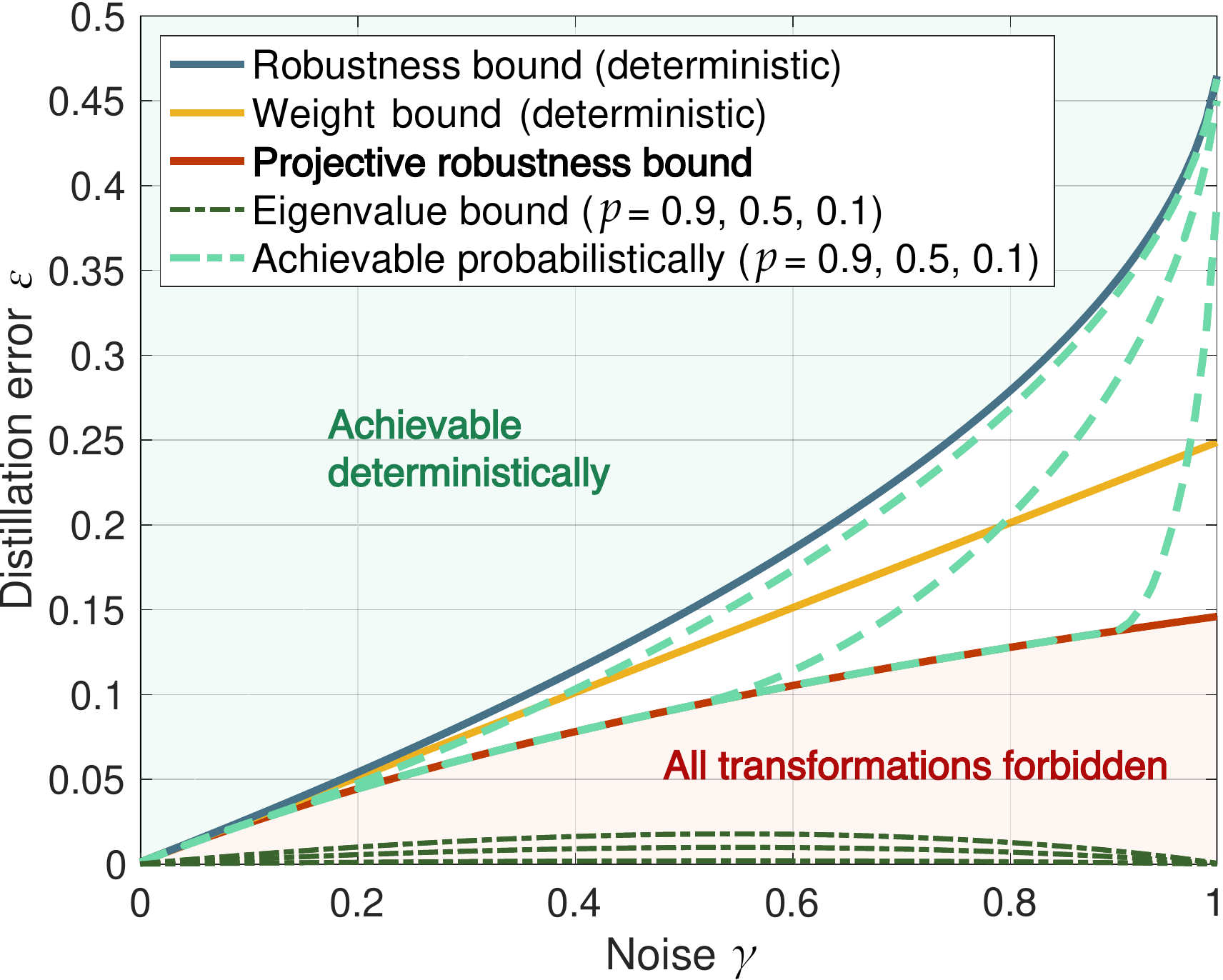}
\caption{\textbf{Bounding achievable error in coherence distillation.} We plot lower bounds on the error $\ve$ incurred in distilling a maximally coherent state $\proj{+}$ from the noisy state $\mathcal{A}_\gamma(\proj{+})$ under maximally incoherent operations~\cite{aberg_2006,chitambar_2016-1}, where $\mathcal{A}_\gamma$ is the amplitude damping channel with damping parameter $\gamma$. Plotted are bounds based on the projective robustness $\RW$ (Theorem~\ref{thm:prob_error}), robustness $R_\FF$~\cite{regula_2017,regula_2020}, weight $W_\FF$~\cite{regula_2021-1,fang_2020-2}, and the eigenvalue bound~\cite{fang_2020}. Achievable performance of probabilistic protocols~\cite{fang_2018} is then plotted for $p \in \{0.9, 0.5, 0.1\}$ (from top to bottom, respectively). 
\\The robustness $R_\FF$ is known to tightly bound deterministic transformations, indicating the region (shaded green) that can be achieved with probability 1~\cite{regula_2017,regula_2020}. On the other hand, we see that $\RW$ tightly bounds the forbidden region, where the given error cannot be achieved with any probability (shaded red). This is indeed the best possible bound, since achievable probabilistic protocols span the entire region between the deterministically achievable error and our probabilistic no-go bound. Conversely, the eigenvalue bound of Ref.~\cite{fang_2020} fails to provide useful restrictions.
}
\label{fig:coh}
\end{figure}

\subsection*{Achievability}
In Figure~\ref{fig:coh} we plot our bound applied to one-shot coherence distillation, showing that Theorem~\ref{thm:prob_error} gives the tightest possible restriction on the achievable error. 
This motivates the question of whether our threshold can exactly characterize probabilistic distillation of resources in more general settings.
Indeed, we show this to be the case whenever the target pure state $\phi$ is chosen to be a maximally resourceful state $\phim$, 
a very natural choice in practical distillation protocols --- for instance, in entanglement theory, this can be understood as a maximally entangled state of some dimension.
\begin{theorem}\label{thm:prob_error_tight}
Let $\phim$ be a state that maximizes the robustness $R_\FF$ among all states of the same dimension. Then, as long as either: (i) the given resource theory is affine, or (ii) it holds that $R_\FF(\phim) = R^\FF_\FF(\phim)$,
then there exists a protocol that achieves the bound of Theorem~\ref{thm:prob_error}. Specifically, for any state $\rho$,
\ba
  \inf_{\E \in \OO} \left[ 1 -  F\left(\frac{\E(\rho)}{\Tr \E(\rho)}, \phim\right)\right] = \left( \frac{F_\FF(\phim)}{1-F_\FF(\phim)} \,\RW(\rho)\, + 1 \right)^{-1}\!\!.
\ea
\end{theorem}
Let us discuss the applicability of this result. In addition to the vast class of all affine resource theories, the Theorem is valid in full-dimensional theories whenever the robustness $R_\FF(\phim)$ equals the standard robustness $R^\FF_\FF(\phim)$, which is satisfied in theories such as entanglement~\cite{harrow_2003} (including entanglement of higher Schmidt rank~\cite{johnston_2018} and genuine multipartite entanglement~\cite{contreras-tejada_2019}) as well as multi-level quantum coherence~\cite{johnston_2018}. Theorem~\ref{thm:prob_error_tight} thus gives an exact expression for the tightest achievable fidelity threshold in resource theories such as coherence, entanglement, or asymmetry.

Theorem~\ref{thm:prob_error_tight} can alternatively be used to quantify the maximal resource that can be distilled from $\rho$ probabilistically, up to given accuracy. For instance, in the resource theory of entanglement with the maximally entangled states $\ket{\phi_m} = \sum_{i=1}^{m} \frac{1}{\sqrt{m}} \ket{ii}$, we have
\ba
  \sup_{\E \in \OO} \lsetr m \barr F\left(\frac{\E(\rho)}{\Tr \E(\rho)}, \phi_m\right) \geq 1-\ve \rsetr = \left\lfloor \frac{\ve}{1-\ve} \RW(\rho) + 1 \right\rfloor\!.\label{eq:distillable_err}
\ea

\subsection*{Many-copy distillation overheads}
\begin{figure}[t!]
\centering
\hspace{-0.05cm}\includegraphics[width=8.5cm]{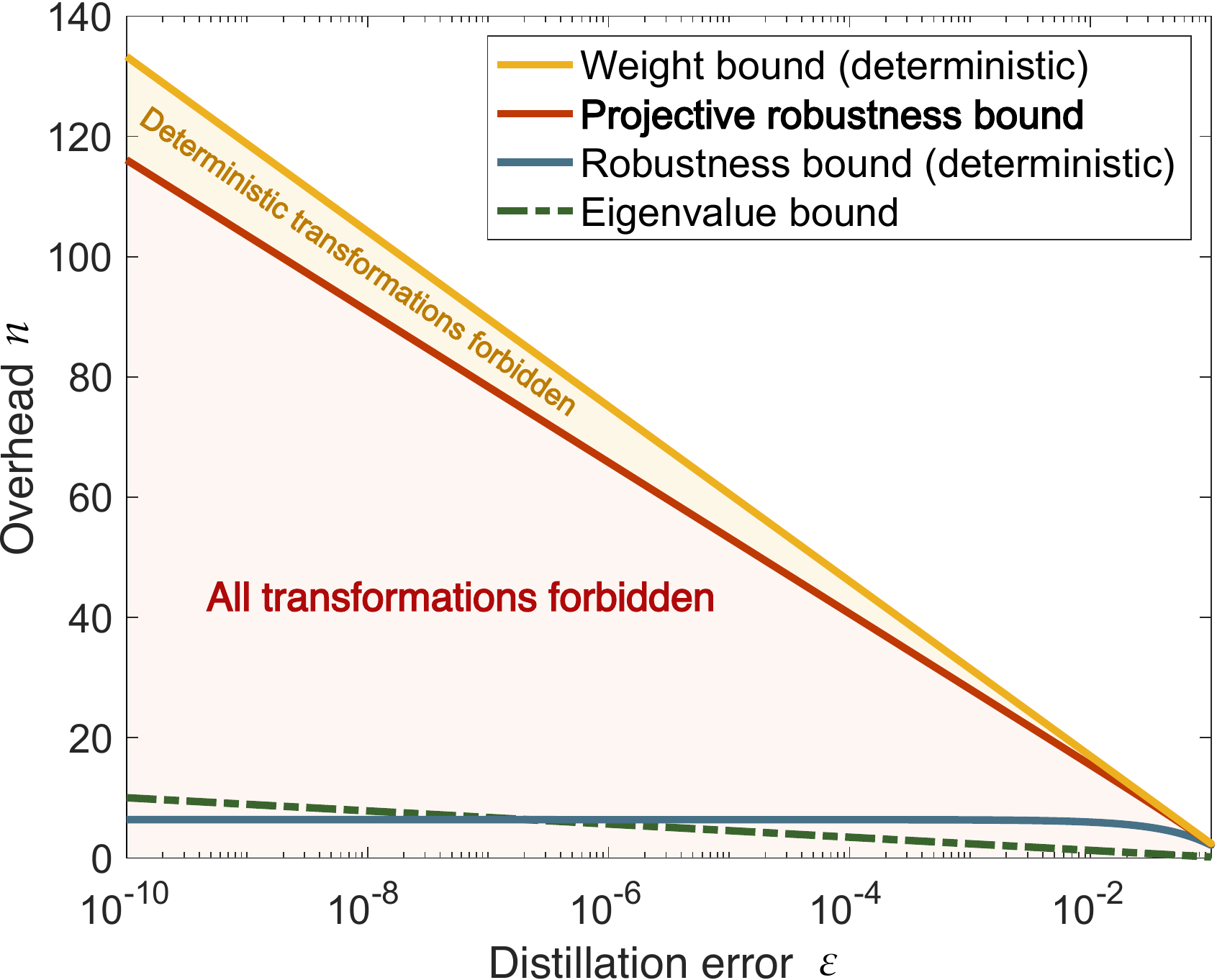}
\caption{\textbf{Bounding probabilistic magic state distillation overhead.} We plot lower bounds on the number of copies of a noisy $T$ state $\rho = \frac{3}{4} \proj{T} + \frac{1}{8} \id$ necessary to distill a noiseless $T$ state $\ket{T} \propto \ket{0} + e^{i \pi /4}\ket{1}$~\cite{bravyi_2005} up to the desired error. 
$\RW$ establishes a lower bound on all transformations, including probabilistic ones, while the bounds based on the weight and the robustness both lower bound the deterministic overhead. The region between the projective robustness bound and the deterministic bounds therefore represents the range where deterministic transformations are not allowed, but probabilistic ones might be possible. We furthermore see that $\RW$ provides a substantial improvement over the eigenvalue bound of Ref.~\cite{fang_2020}.}

\label{fig:magic}
\end{figure}

In practice, one often aims to minimize the transformation error by using more input copies of the given state. 
An important application of Theorem~\ref{thm:prob_error} is to lower bound the overhead required in many-copy transformations.
\begin{corollary}\label{cor:overhead}
Assume that the given resource theory is closed under tensor product, i.e.\ $\sigma \in \FF \Rightarrow \sigma \otimes \sigma \in \FF$.

Then, any free probabilistic transformation $\rho^{\otimes n}~\to~\tau$ such that $\tau$ is a state satisfying $F(\tau,\phi) \geq 1-\ve$ for some resourceful pure state $\phi$ requires at least 
\begin{equation}\begin{aligned}
     n \geq \log_{\RW(\rho)} \frac{(1-\ve)\left[1-F_\FF(\phi)\right]}{\ve\, F_\FF(\phi)}
 \end{aligned}\end{equation} copies of $\rho$.
\end{corollary}
We demonstrate the performance of this bound in magic state distillation in Figure~\ref{fig:magic}, where we see that it is not much lower than the best known deterministic bound given by the weight $W_\FF$~\cite{regula_2021-1,fang_2020-2}.
This could then suggest that employing probabilistic manipulation schemes cannot give significant advantages over deterministic ones in practical error regimes. However, a better understanding of how closely the bounds can be approached by feasible distillation protocols would be required to make definitive conclusions.

\subsection*{Conclusions}
We introduced the projective robustness $\RW$, a powerful resource monotone that allowed us to reveal universal restrictions on the manipulation of quantum resources.
We established no-go theorems for probabilistic transformations of quantum states in arbitrary resource theories, which in fact become necessary and sufficient for certain types of resources such as all affine theories. We demonstrated the usefulness of the restrictions by applying them to the problem of distillation, establishing bounds that conclusively rule out the possibility of purifying resources in certain error regimes.

Beyond general bounds on the capabilities of all probabilistic protocols as established here, one might be interested in better understanding the achievable performance of resource transformations with some probability, and in particular the possible trade-offs between probabilities and transformation errors. We address this in~\cite{regula_2021-4} with a complementary approach, also based on convex optimization.

$\,$

\begin{acknowledgments}
I would like to thank Alexander Streltsov and Ryuji Takagi for useful discussions. This work was supported by the Japan Society for the Promotion of Science (JSPS) KAKENHI Grant No.\ 21F21015 and the JSPS Postdoctoral Fellowship for Research in Japan.
\end{acknowledgments}

 \bibliographystyle{apsc}
 \bibliography{../main}

\end{document}